\documentclass[twoside,twocolumn,10pt]{article}

\usepackage{fancyhdr}

\makeatletter
\newsavebox{\@linebox}
 \savebox{\@linebox}[3em][t]{\parbox[t]{3em}{%
   \@tempcnta\@ne\relax
   \loop{\underline{
     \the\@tempcnta}}\\
     \advance\@tempcnta by \@ne\ifnum\@tempcnta<57\repeat}}

\fancypagestyle{numberstyle}{
 \fancyhead{}
 \fancyfoot{}
 \fancyhead[LE]{
     \begin{picture}(0,0)%
      \put(-20,-15){\usebox{\@linebox}}%
      \put(480,-15){\usebox{\@linebox}}%
      \end{picture}}

 \fancyhead[LO]{
    \begin{picture}(0,0)%
      \put(-30,-15){\usebox{\@linebox}}%
      \put(470,-15){\usebox{\@linebox}}%
     \end{picture}}
\fancyfoot[C]{\scriptsize \thepage}
\makeatother
}
\usepackage[pscoord]{eso-pic}
\usepackage[color=black,opacity=1,angle=0,scale=1]{background}
\newcommand{\placetextbox}[3]{
  \setbox0=\hbox{#3}
  \AddToShipoutPictureFG*{
    \put(\LenToUnit{#1\paperwidth},\LenToUnit{#2\paperheight}){\vtop{{\null}\makebox[0pt][c]{#3}}}%
  }%
}%

\backgroundsetup{
  contents={%
  \placetextbox{0.5}{.95}{Accepted Manuscript at WSCG 2021}
  \placetextbox{0.5}{0.085}{p. \thepage}%
  }
}

\usepackage[sorting=none]{biblatex} 
\bibliography{report.bib}

\usepackage{subfigure}

\usepackage{wscg}           
\RequirePackage{ifpdf}
\ifpdf
\else
 \RequirePackage[dvips,draft]{graphicx}
 \RequirePackage[dvips]{color}
\fi

\usepackage{nopageno}       
\usepackage{placeins}

\title{Comparison of 2D vs. 3D U-Net Organ Segmentation in abdominal 3D CT images}

\author{
\hspace{-1cm}
\parbox{0.35\textwidth}{\centering
Nico Zettler\\
Aalen University\\
Beethovenstr. 1\\
Burren Campus\\
Germany, 73430, Aalen, Baden-Wuerttemberg\\[1mm]
nico.zettler@hs-aalen.de
}
\parbox{0.35\textwidth}{\centering
Andre Mastmeyer\\
Aalen University\\
Beethovenstr. 1\\
Burren Campus\\
Germany, 73430, Aalen, Baden-Wuerttemberg\\[1mm]
andre.mastmeyer@hs-aalen.de
}
}

\usepackage{url}
\makeatletter
\def\Uslash{\mathbin{\mathchar`\/}\@ifnextchar{/}{\kern-.15em}{}}
\g@addto@macro\UrlSpecials{\do \/ {\Uslash}}
\def\Ucolon{\mathbin{\mathchar`:}\@ifnextchar{/}{\kern-.1em}{}}
\g@addto@macro\UrlSpecials{\do : {\Ucolon}}
\makeatother

\begin{document}

\twocolumn[{\csname @twocolumnfalse\endcsname

\maketitle  

\begin{abstract}
\noindent
A two-step concept for 3D segmentation on 5 abdominal organs inside volumetric CT images is presented. First each relevant organ's volume of interest is extracted as bounding box. The extracted volume acts as input for a second stage, wherein two compared U-Nets with different architectural dimensions re-construct an organ segmentation as label mask. In this work, we focus on comparing 2D U-Nets vs. 3D U-Net counterparts. Our initial results indicate Dice improvements of about 6\% at maximum. In this study to our surprise, liver and kidneys for instance were tackled significantly better using the faster and GPU-memory saving 2D U-Nets. For other abdominal key organs, there were no significant differences, but we observe highly significant advantages for the 2D U-Net in terms of GPU computational efforts for all organs under study.
\end{abstract}

\subsection*{Keywords}
Organ bounds, U-Net, Architecture, Abdomen, Segmentation, 3D CT images
\vspace*{1.0\baselineskip}
}]


\section{Introduction}

\copyrightspace

Manual segmentation of liver, kidneys, spleen and the low-contrast pancreas in axial CT slices is very time consuming for radiologists. The automated just-in-time (JIT) segmentation of 3D patient organ models continues to be an unsolved challenge. Automatic and JIT reconstruction procedures of 3D models are highly relevant for surgery planning and navigation. Another important aspect is the saving of computational power and memory, which differ greatly among various Convolutional Neural Network (CNN) architectures for image segmentation. The 2D U-Net architecture\cite{Ronneberger15} showed promising results in 2015, outperforming conventional models on 2D biomedical segmentation problems and being effective on smaller images. Unlike previously known CNN models, the U-Net concept uses down- and up-sampling steps to resample the condensed feature map to the original size of the input image. By incorporating higher resolution feature information in each upsampling step, semantic segmentation of the input images can be achieved very efficiently by skip connections. The resulting U-shaped architecture was extended to a 3D U-Net edition in 2016 by Cicek et al.\cite{Cicek16} by replacing 2D operations with their 3D equivalents. The research focus of this paper targets the question, is the 3D U-Net really better suited for 3D data?

In Big Data studies we expect, parallel segmentation of thousands of 3D volumetric images requires high computational time due to the limited amount of processing nodes and sub-processes for parallel processing. We suppose, 3D U-Nets demand even higher GPU computational and memory overhead for 3D image processing. In this work, the focus lies on proposing solutions with costs as low as possible while keeping or even improving segmentation quality. In our concept, we (1) detect and let U-Nets work in local sub-images for each task (liver, kidneys, spleen, pancreas). We (2) compare two U-Net architectures for quality and GPU-performance carrying out semantic segmentations. (3) Based on a detailed statistical analysis, we recommend the more suitable architectural model to the interested reader. Again, subject to our image data base of 80, could the simpler 2D U-Net perform better than complex and theoretically more powerful 3D U-Nets?

\section{Recent solutions}

3D segmentation is especially relevant for the emerging field of intervention training and planning using Virtual Reality (VR) techniques. Time-variant 4D VR simulations with breathing simulation are readily available for training and planning of liver interventions\cite{mastmeyer2018population,Mastmeyer2017inter,Mastmeyer2016RTUS}. The JIT and high quality reconstruction of all the necessary 3D models remains a main hindrance, i.e. especially abdominal organs are rarely easy to segment. This is due to varying imaging conditions such as contrast agent administration, structure variations and noise.

To this aim in the literature, 3D U-Nets are very often proposed as mightier\cite{du2020medical,siddique2020u,radiuk2020applying,tajbakhsh2020embracing,litjens2017survey}.
However, there are some studies indicating benefits of 2D U-nets in 3D segmentation tasks such as\cite{nemoto2020efficacy,meine2018comparison,ushinsky20213d}. Nemoto et al.\cite{nemoto2020efficacy} state that low computational burden 2D U-Nets are effective and on par with 3D U-Nets for semantic lung segmentation excluding the trachea and bronchi.

Chang et al.\cite{Chang2018} suggested a hybrid approach of 3D and 2D inputs for the evaluation of hemorrhage on head CT scans, which was later adapted by Ushinsky et al.\cite{ushinsky20213d} for application to prostate segmentation in MRI images, showing that 2D U-nets are very effective for 3D data.
Christ et al.\cite{Christ_2016} demonstrated a slice-wise application of 2D U-Nets to liver segmentation in combination with 3D random fields.
Similarly Meine et al.\cite{meine2018comparison} proposed liver segmentation methods with the assistance of 3D, 2D and three fused 2D U-Net sectional (axial, coronal, sagittal) results in a 2.5D ensemble approach. They find the 2.5D U-Net ensemble results statistically superior for liver segmentation, especially for images with pathologies.

Other recent approaches aim to combine multiple stacked 2D U-Nets\cite{Jha2020} and further improve information flow through semantic connections between different components\cite{Tang2018}. This concept has also been used in application areas such as colon polyp segmentation\cite{sang2021agcuresnest} and face recognition\cite{Na_2020}. Zwettler et al.\cite{zwettler2020} have recently shown that extending datasets with synthesized slices can particularly improve the results of 2D U-Nets applied on a small number of training datasets, indicating futher potential for the 2D approach.

In our 2D U-Net setup for the abdominal organs liver, spleen, kidneys and pancreas, we favor axial section training inside of pre-detected organ-specific VOIs. This is for significant radiation dosage savings by high scan pitch. Yet image resolution in axial CT slices is very high. New aspects in this work are the organ-specific VOI approach for the organs under study such as liver, spleen, kidneys and pancreas. Concluding, an organ-specific architecture dimensionality recommendation for each of the organs is given.

\section{Proposed solutions}

Eighty CT-scans and related label data found in various public sources\footnote{http://visceral.eu, \\http://sliver07.org,
\\http://competitions.codalab.org/competitions/17094} were considered for training and tests of our U-Nets. Their image information differs not only in quality, noise and field of view. But also, the patient-individual volume appearance varies in the amount of slices (64 to 861), pathological lesions, slice width (1 to 5 mm) and applied contrast agent.  

The issue relative to data annotation was mainly the lowly contrasted pancreas organ in the CTs, unavailable with some image sets consisting of congruent intensity and label images from the public sources. We coped with that by four eye reviewed manual segmentation of this occasionally missing structure in the label maps. However, most organ reference segmentation were readily available in the public data sources.

\subsection*{Preprocessing}

The orientation of images was changed to Right-Anterior-Inferior (RAI) and zero origin (0.0, 0.0, 0.0).
As CNNs are not able to interpret voxel spacings natively, isotropic
image resampling  with 2.0$^3$ mm as trade-off for varying image xyz-spacings (xy $\le$ 2 mm, z $\le$ 5 mm) was performed.

Our concept is composed of two different machine learning approaches, bounding box detection using random regression forests (RRF) and U-Nets for semantic segmentation. 
The RRFs can be used to detect organ VOIs in CT data. The U-Nets use the detected VOIs and segment the contained organs.

\subsection*{Volume of interest bounding box detection}

The ground truth corresponding organ VOI bounding box (BB) - needed in this work's evaluation - of each of the organs can be created by scanning the reference segmentation maps for labeled voxel coordinate extremes. To create a three-dimensional BB vector for each organ, for each coordinate direction \((x,y,z)\), we iterate in orthogonal slices through the organ's label map and save the extreme limits in a 6D BB vector to serve as ground truth BBs.

Alternatively, VOI-based organ extraction\cite{kern20203d,Mastmeyer2016RF,Criminisi13,} is readily available for organ-specific VOI detection of the organ ensemble (liver, still left kidney, right kidney, spleen, pancreas). This step also solves the FoV problem in the scanner z-direction, as CT scans cover variable body portions. Our currently investigated scheme to learn bounding boxes of VOIs using RRFs is summarized in the following. 

\begin{figure}[!h]
\centering
\includegraphics[width=5cm]{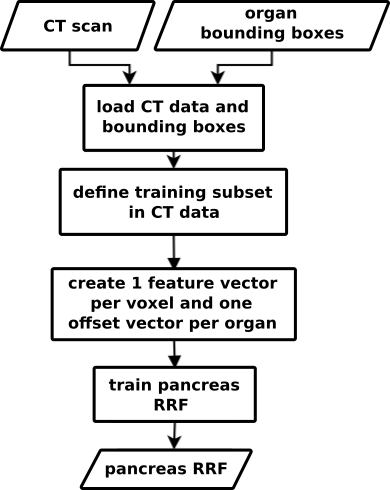}
\caption{The inputs for the training process are CT scans and ground truth VOIs of a targeted pancreas. We create one feature vector and one mm offset vector for each voxel that is part of a predefined medial cylinder subset in the scan\cite{Criminisi13}. The trained RRF is able to predict the offset between a voxel and an organ's VOI walls.}
\label{fig:1}
\end{figure}

In our concept, we use Random Regression Forests (RRF) to determine the location and extent of abdominal organs\cite{Criminisi13}. As seen in Fig. \ref{fig:1}, the RRF training step expects scans and ground truth VOIs as input.

A three-dimensional VOI \(b_c\) of an organ \(c\) can be described by using a 6D vector \(b_c=(b^{Left}_c, b^{Right}_c, b^{Anterior}_c, b^{Posterior}_c, b^{Head}_c, b^{Foot}_c )\) with coordinates in mm\cite{Criminisi13}. We run over all voxels \(p = (x_p,y_p,z_p)\), which are within a specified radial distance (r = 5\,cm) from the scan medial axis. The distance \(d\) between such a voxel and each of the VOI walls, can be calculated by using \(d(p)= (x_p-x^{Left}, x_p-x^{Right}, y_p-y^{Anterior}, y_p-y^{Posterior}, z_p-z^{Head}, z_p-z^{Foot})\) and is saved as the offset-vector to be learned. In contrast to Criminisi et al.\cite{Criminisi13}, we use only 50 feature boxes, that are evenly distributed on three spheres (r = 5\,cm, 2.5\,cm, 1.25\,cm) to generate the input feature vector. The feature boxes \(F_j\) are intended to capture the spatial and intensity context of the current voxel. For this purpose, the mean intensities of the feature boxes are calculated and saved in the feature vector.

Within the first step of our application concept (Fig. \ref {fig:frameworkScheme}, left) for every single organ, the RRF locates the VOI for a particular organ as BB. 

A detected VOI's information (intensities, labels) is finally resampled to 96x96(x96) as CNN input.

\subsection*{Training and application of  2D and 3D U-Net architectures}

\begin{figure*}[hbt!]
 \centering
 \frame{\includegraphics[trim={0cm 2cm 0.5cm 0.5cm},clip, scale=0.525]{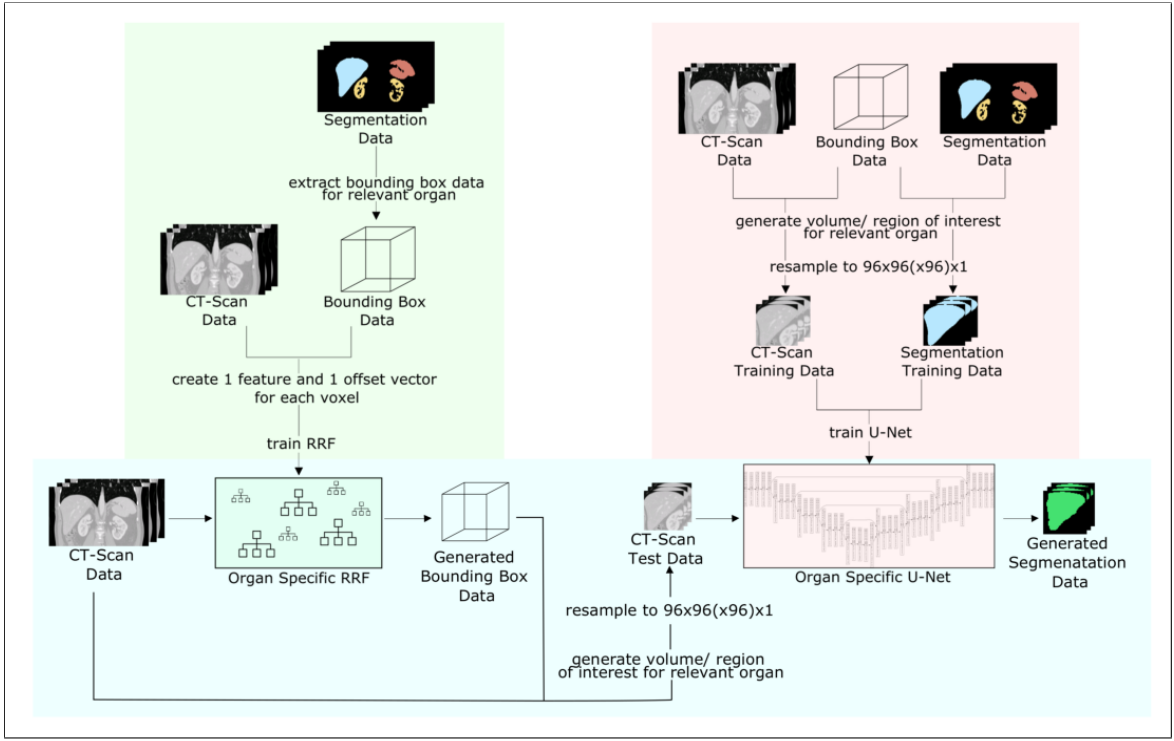}}
 \caption{Bird's eye perspective on our current segmentation concept with RRF for VOI detection and subsequent U-Net. Both training (green, red) and application (blue) concepts are shown. In this work, the right part with regards to the U-Nets is focused in the evaluation}
 \label{fig:frameworkScheme}
\end{figure*}

Within the 2nd stage, as shown in the bird's eye view in Fig. \ref{fig:frameworkScheme}, right, the obtained VOI is either passed to a 2D\cite{Ronneberger15} or a 3D U-Net\cite{Cicek16} for semantic labeling.

\begin{figure}[!h]
\centering
\includegraphics[width=5cm]{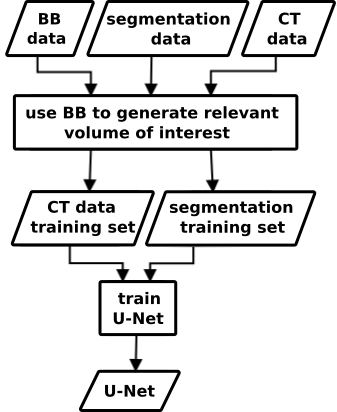}
\caption{The inputs for the training process are ground truth bounding boxes VOIs, CT scans and their corresponding segmentation maps. The box cuts out the CT- and segmentation data to extract the relevant image region. Inside the organ VOIs, the segmentation is learned. The process results in organ-wise U-Nets}
\label{fig:3}
\end{figure}

The training data for our U-Net consists of the expert segmentations and ground truth bounding box contents, as schematized in Figs. \ref{fig:3}, \ref{fig:frameworkScheme}. The VOIs are then used to locally extract the intensity and label data from the CT scans. As input, a U-Net receives a VOI from the intensity data, while the corresponding label data is connected to the output. We use lightly modified 2D and 3D U-Net architectures vs. Ronneberger et al.\cite{Ronneberger15}. Our architecture consists of four down- and up-scaling steps connected by skip connections and max-pooling while down-sampling.
For the 3D U-Net, an additional layer depth dimension is added while all other design elements are kept constant.

The U-Net uses the data contained inside a given VOI bounding box to segment the corresponding organ. The output is a segmentation map of the full target organ in a probability range from 0 to 100\%. The threshold for all organs except pancreas was chosen as 50\%. For pancreas 30\% was empirically found best.

The U-Nets were trained using batches of size 8 over 100 epochs. In addition, Adam optimization and a cross entropy loss function were used. We train one U-Net for each organ, using a ReLU activation function.

\subsection*{Evaluation, metrics and statistics}

For fair comparison, in each cross-validation iteration, we use the same 64 training images for both U-Nets, and test with the same 16. This way, we prevent data contamination by separate training and test sets. The difference between the U-Net architectures is input, output and filter kernel layer dimension only. For 2D U-Nets, input and output size is 96x96 and 2D filter kernels are used in the layers. Accordingly for 3D U-Nets, we have 96x96x96 in- and outputs and use 3D convolutions in the architecture layers. 

A 5-fold randomized cross-validation using 4:1 splits was used. 
Five iterations yield 80 quality measures for each organ being used in our statistics (Tab. \ref{tab:dice}). This means five new models for each organ are trained with randomly selected training data and used in the evaluation. To purely concentrate on the influence associated with U-Net dimensionality, the reference VOIs were utilized in the evaluation of this study. The metric used for analysis is the Dice similarity coefficient:
\begin{equation}
    DSC=\frac{2\cdot|U \cap G|}{|U| + |G|},
\end{equation}
where $U$ represents the voxel set from U-Net object segmentation and $G$ the ground truth voxels. A DSC value of 1 indicates perfect segmentation, a value of near 0 poor segmentation.
From the DSC results, we calculate means and standard deviations, medians and Inter-Quartile-Ranges (IQR) as measures of accuracy and precision.
Statistical assessments using paired T-tests and Wilcoxon-Signed-Rank(WSR)-test were performed with GNU-R 4.0.3. We finally recommend the dimensionality of the U-Net architecture based on accuracy, i.e. greater mean or median, and precision, i.e. smaller standard deviation or IQR and smaller number of outliers. We also keep an eye on the ratio between quality and GPU efforts in time and memory.

\section{Experimental results}

\begin{figure*}
\centering
\includegraphics[width=0.75\textwidth]{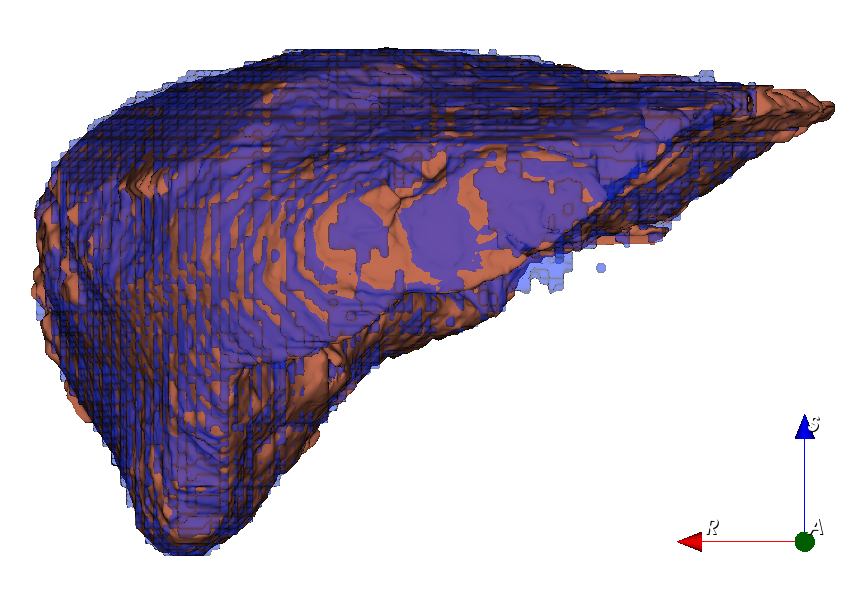}\\
\includegraphics[width=0.75\textwidth]{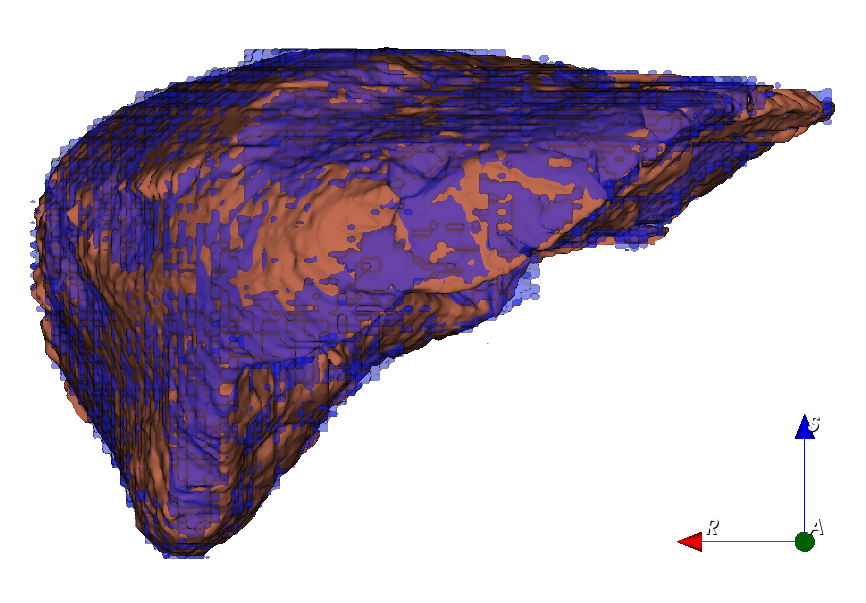}
{%
\caption{\label{fig:images}Two liver segmentations (anterior view) with 2D U-Net (top) and 3D U-Net (bottom). The coverage of the 2D U-Net result (top) appears more equally distributed, i.e. more sensitive. Brown: Reference and purple: U-Net CNN. N.B., for the 2D model result on top, the oscillation pattern between reference (brown) and 2D CNN segmentation (purple) is much denser, thus showing out better mean local quality.\\\textbf{Legend}: reference (brown) and U-Net segmentation (purple)}
}
\end{figure*}

Qualitative liver results (Fig. \ref{fig:images}) show the 2D U-Net (top) superior with a more evenly distributed segmentation surface coverage. The 3D U-Net (bottom) obviously suffers from under-segmentation as more brown ground truth surface is visible in Fig. \ref{fig:images} (bottom). Referring to Tab. \ref{tab:dice}, precision is also better for the 2D U-Net due to lower std. deviations and IQRs.

\begin{figure*}[hbt!]
    \centering
    \subfigure[Liver]{\label{plot:boxplotOverviewLiver}
    {\includegraphics*[width=7.25cm]{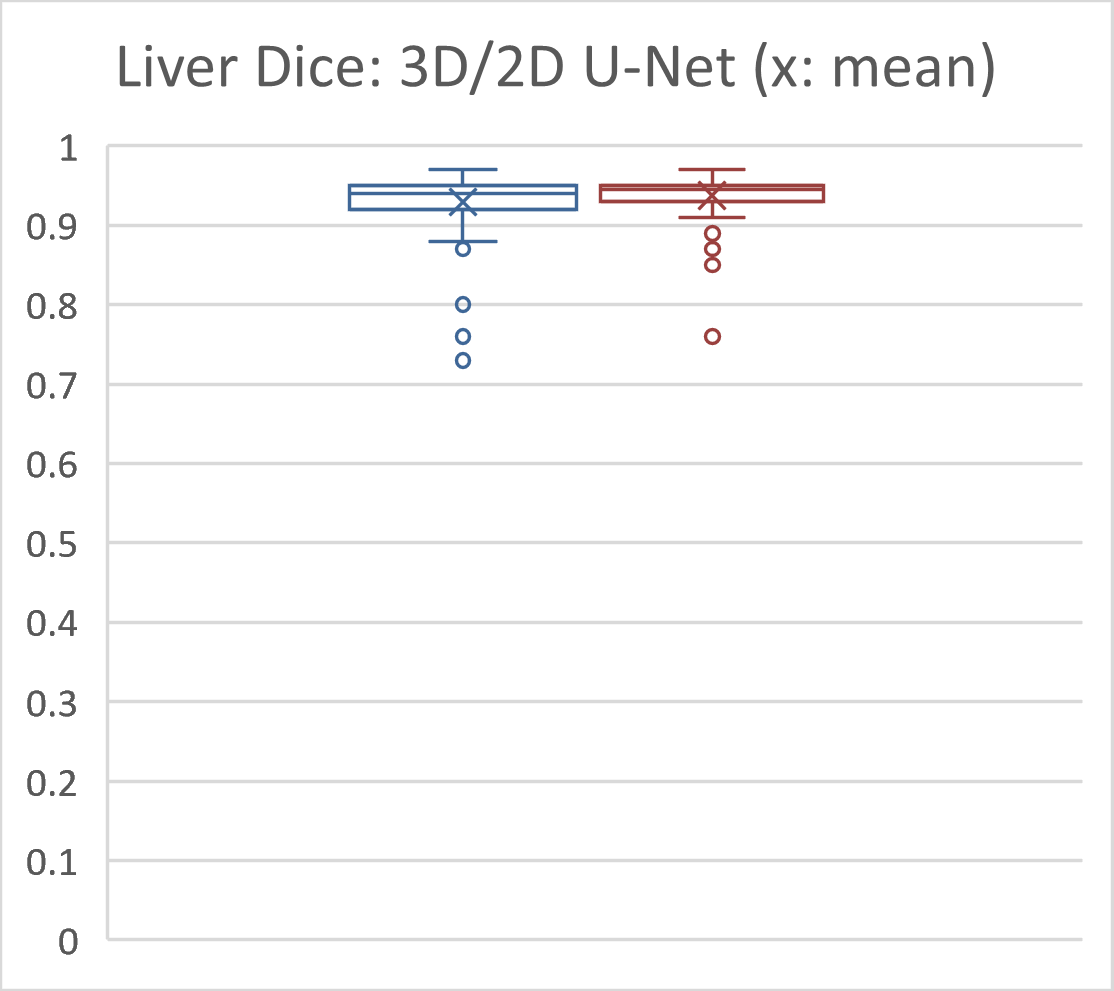} 
    }}%
    \qquad
    \subfigure[Left and right kidneys]{\label{plot:boxplotOverviewKidneys}
    {\includegraphics*[width=7.25cm]{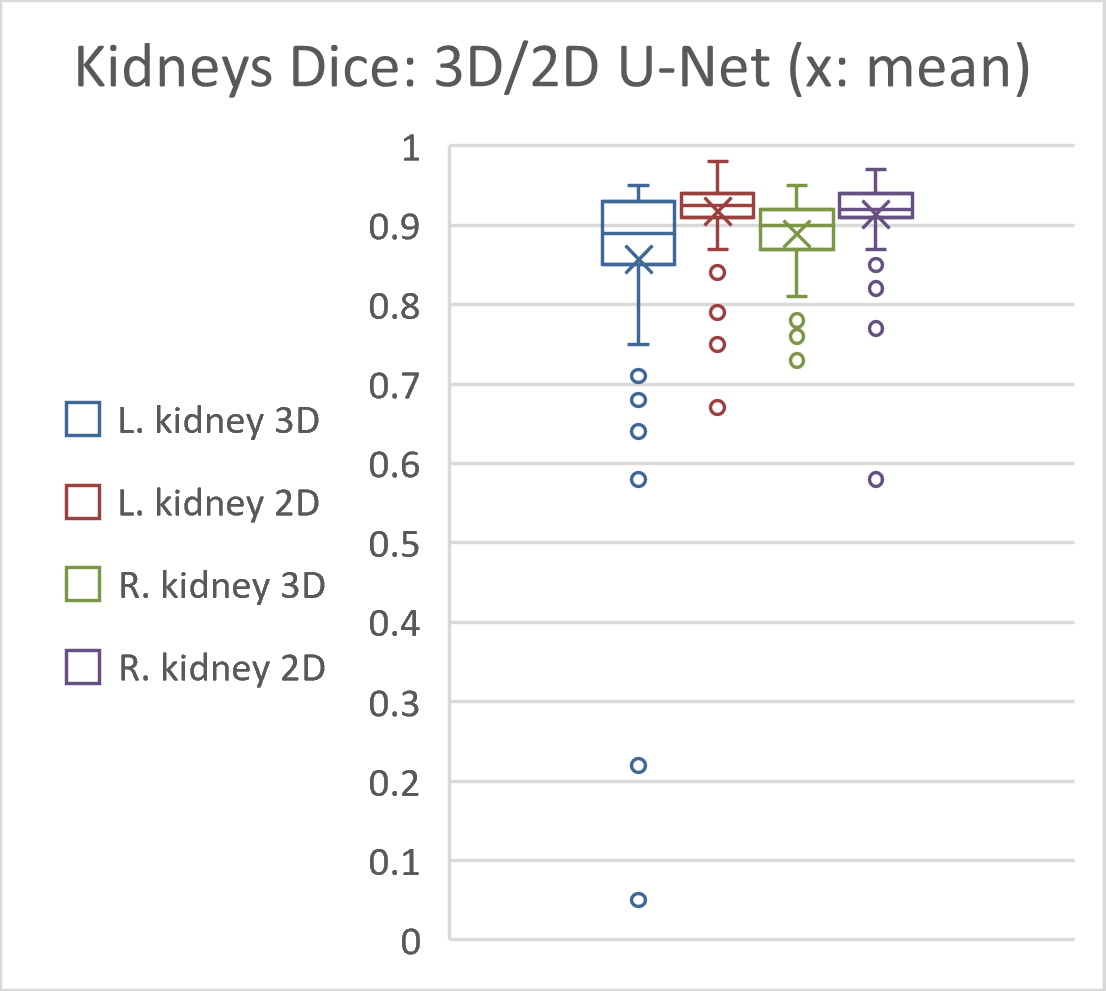} 
    }}%
    \vspace{-.15cm}
    \qquad
    \subfigure[Spleen]{\label{plot:boxplotOverviewSpleen}
    {\includegraphics*[width=7.25cm]{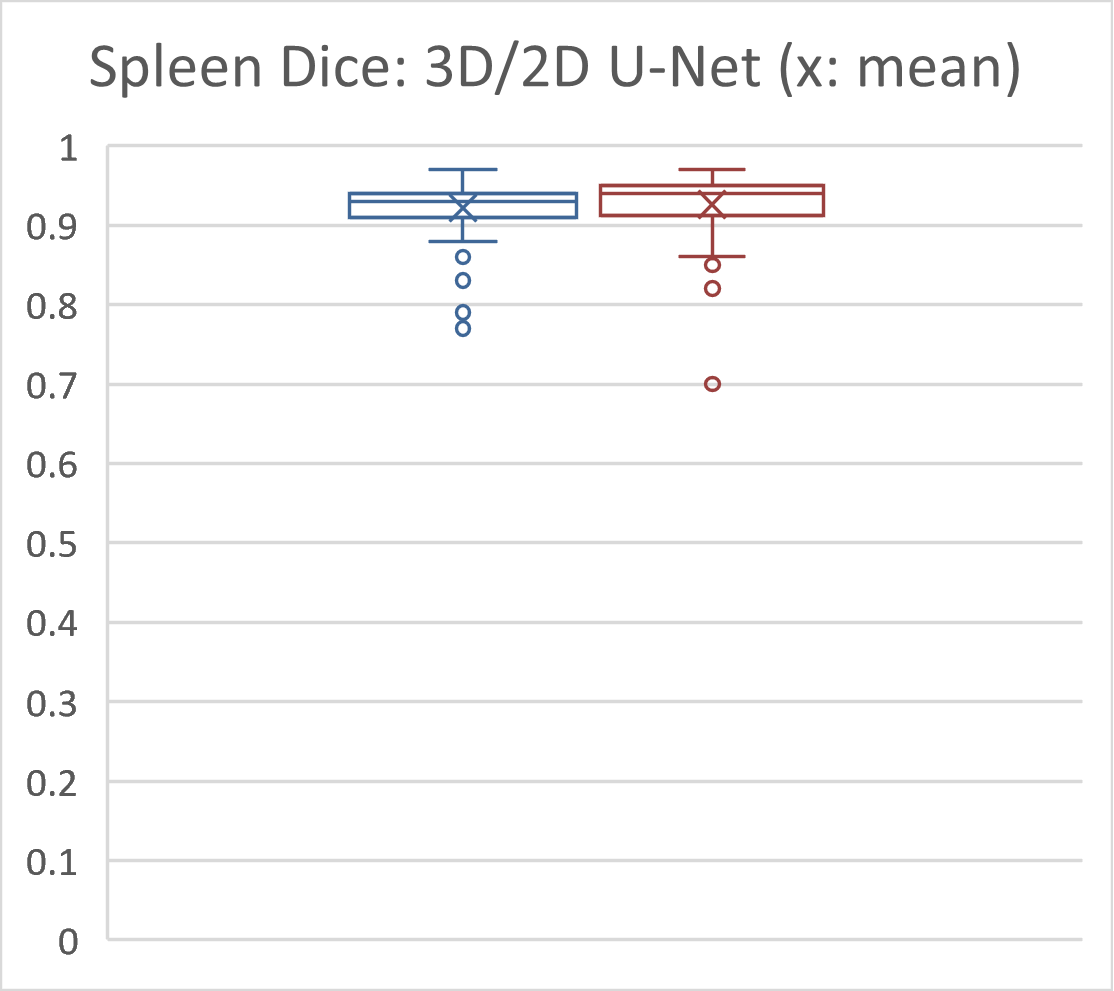} 
    }}%
    \qquad
    \subfigure[Pancreas]{\label{plot:boxplotOverviewPancreas}
    {\includegraphics*[width=7.25cm]{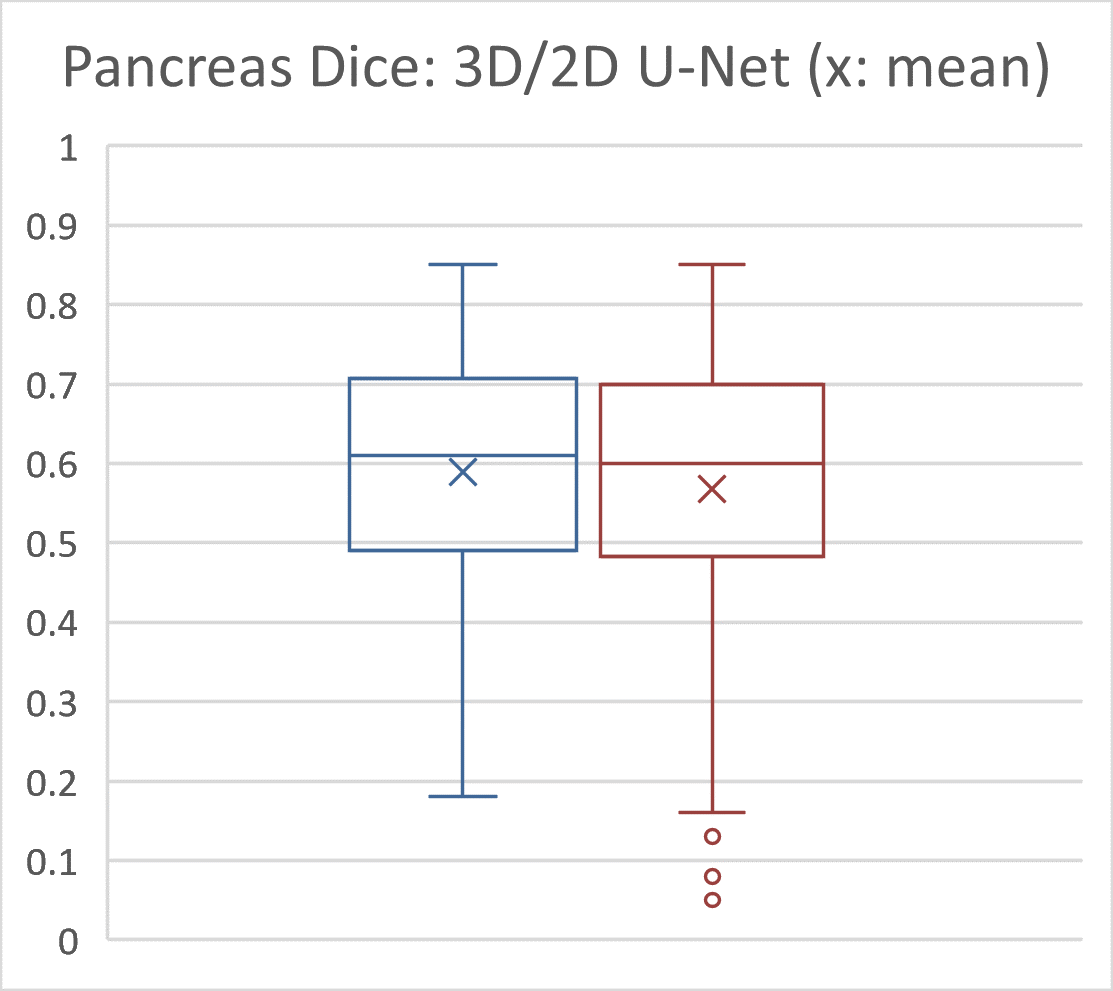} 
    }}%
    \vspace{-.15cm}
    \caption{Overview DSC boxplots: 3D (blue, green) and 2D U-Net (brown, purple) on the x-axis vs. DSCs on the y-axis: 2D U-Nets in favor for (a) liver and (b) kidneys (left k. and 3D U-Net: blue, left k. and 2D U-Net: brown; right k. and 3D U-Net: green,  right k. and 2D U-Net: purple). (c) Spleen comes out with an edge for the 2D U-Net by trend. (d) Mixed results: 2D U-Net wins the accuracy contest, but loses the precision contest in terms of lower standard deviation (cf. Tab. \ref{tab:dice}) and regarding less outliers for pancreas.\\\textbf{Legend} for (a), (c) and (d): blue boxplots: 3D U-Net; red boxplots: 2D U-Net.}%
    \label{fig:overviewBoxplots}%
\end{figure*}

\begin{figure*}[hbt!]
    \centering
    \subfigure[Liver]{\label{plot:boxplotLiver}
    {\includegraphics*[width=7.25cm]{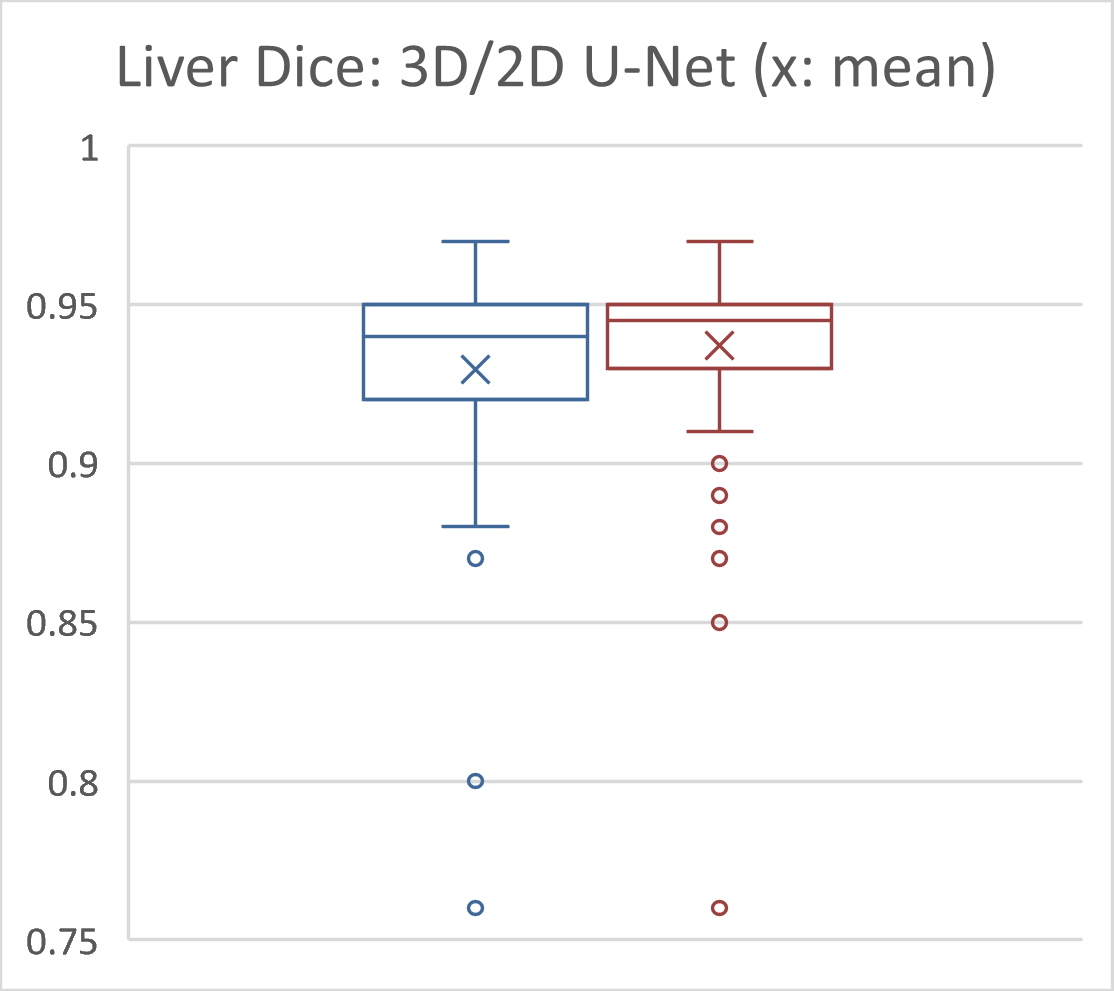} 
    }}%
    \qquad
    \subfigure[Left and right kidneys]{\label{plot:boxplotKidneys}
    {\includegraphics*[width=7.25cm]{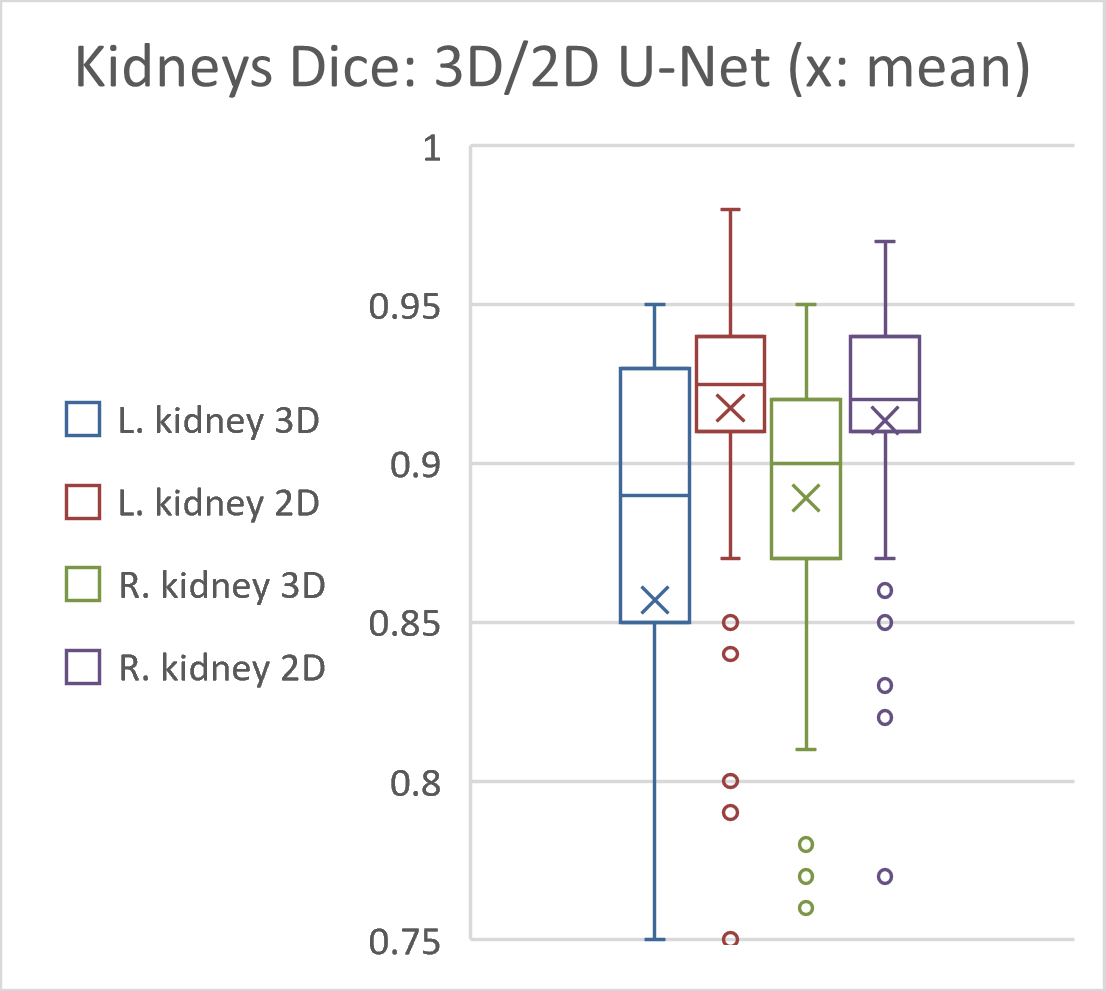} 
    }}%
    \vspace{-.15cm}
    \qquad
    \subfigure[Spleen]{\label{plot:boxplotSpleen}
    {\includegraphics*[width=7.25cm]{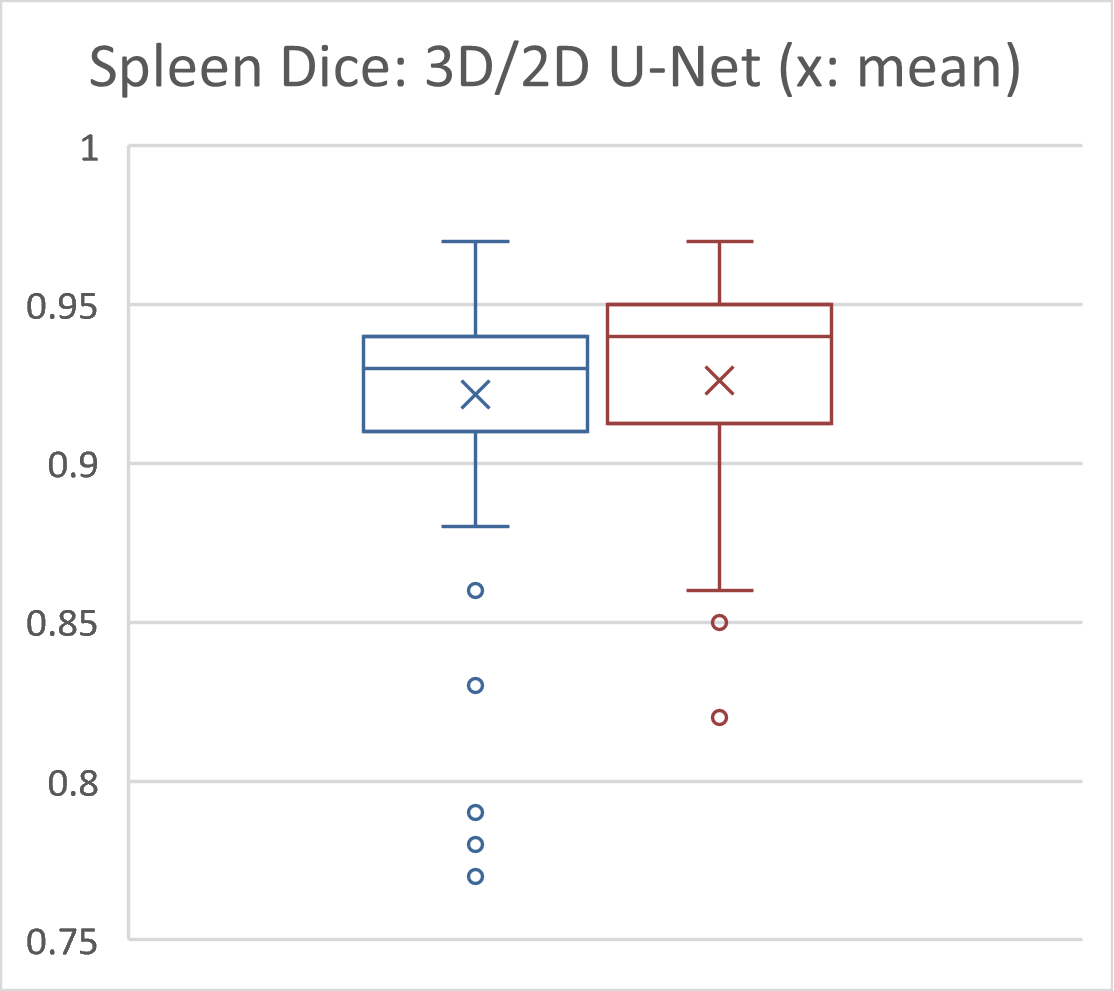} 
    }}%
    \qquad
    \subfigure[Pancreas]{\label{plot:boxplotPancreas}
    {\includegraphics*[width=7.25cm]{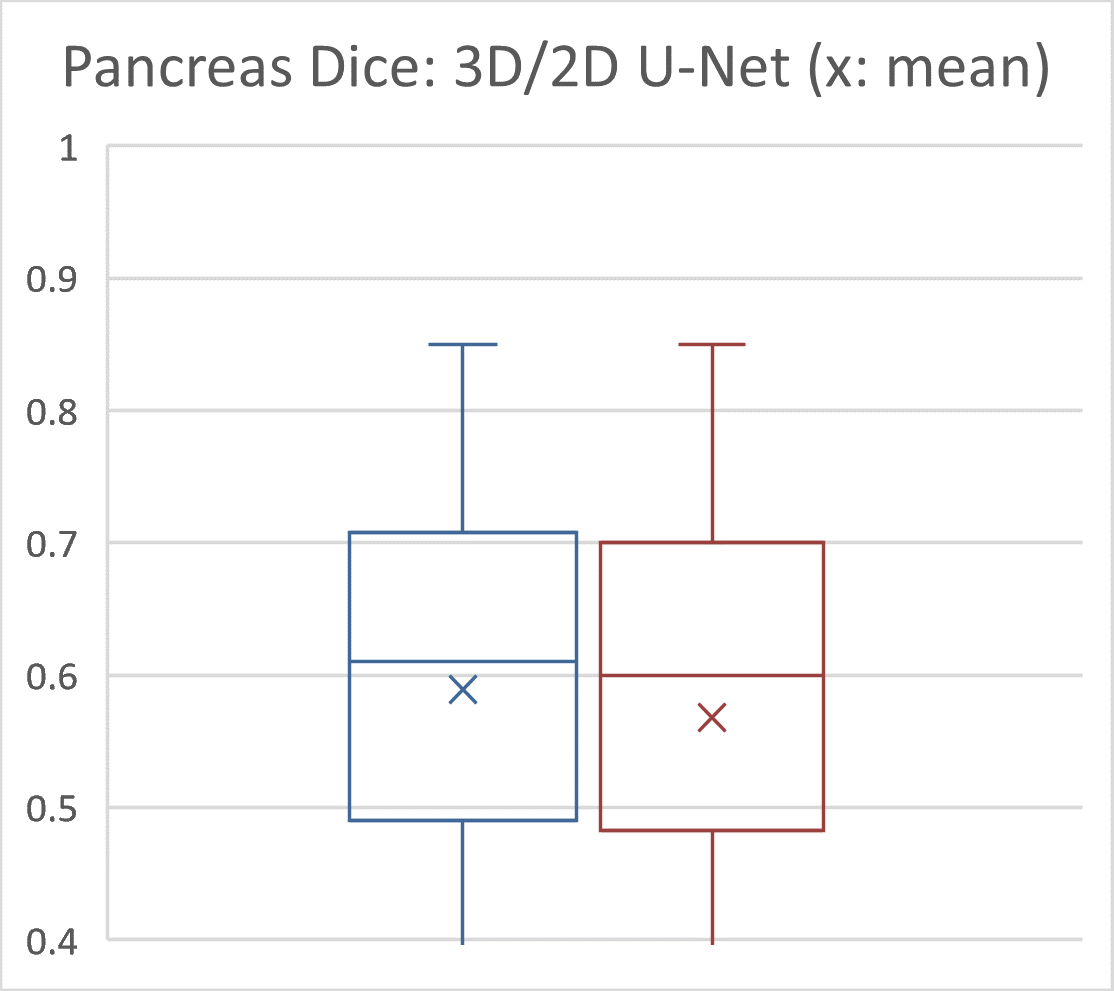} 
    }}%
    \vspace{-.15cm}
    \caption{Zoomed DSC boxplots: 3D (blue, green) and 2D U-Net (brown, purple) on the x-axis vs. DSCs on the y-axis: 2D U-Nets in favor for (a) liver and (b) kidneys (left k. and 3D U-Net: blue, left k. and 2D U-Net: brown; right k. and 3D U-Net: green,  right k. and 2D U-Net: purple). (c) Spleen comes out with an edge for the 2D U-Net by trend. (d) Mixed results: 2D U-Net wins the accuracy contest, but loses the precision contest in terms of lower standard deviation (cf. Tab. \ref{tab:dice}) and regarding less outliers for pancreas.\\\textbf{Legend} for (a), (c) and (d): blue boxplots: 3D U-Net; red boxplots: 2D U-Net.}%
    \label{fig:boxplots}%
\end{figure*}

For kidneys, the 2D U-Net is highly significantly better in both T- and WSR-tests. The boxplots with a small x for the mean in Figs. \ref{fig:overviewBoxplots}, \ref{fig:boxplots} confirm visually: the 2D U-Net is significantly better for liver and highly significantly better for kidneys (Figs. \ref{plot:boxplotOverviewLiver}, \ref{plot:boxplotLiver}, \ref{plot:boxplotOverviewKidneys}, \ref{plot:boxplotKidneys}).

Spleen segmentation shows an advantage for the 2D U-Net by trend as shown in Figs. \ref{plot:boxplotOverviewSpleen}, \ref{plot:boxplotSpleen}) and Tab. \ref{tab:dice}. 

The 3D U-Net possesses higher accuracy and precision for the low-contrast pancreas. Higher precision is indicated by smaller standard deviations and less outliers in Tab. \ref{tab:dice} and Figs. \ref{plot:boxplotOverviewPancreas} and \ref{plot:boxplotPancreas}. With regards to accuracy measured by medians, the 3D U-Net has an edge by trend of 0.02 over the 2D U-Net, while precision measured by IQR is the same.

The GPU-Performance evaluation in Tab. \ref{tab:GPU-performance} shows the superiority of the 2D U-Net in this study's scale with an 80 images data base. The ratio of quality to GPU resources is always better for the 2D U-Net. In Tab. \ref{tab:GPU-performance} GPU memory saving is >6-fold in the training stage and >5-fold in the model application, trivially a highly significant result as there are no variations, i.e. no standard deviation and IQR.
With respect to GPU calculation times measured in training, the 2D U-Net is precisely 40 seconds or 7\% faster on average, again a highly significant mean result. In the U-Net model application, we can observe a weakly significant ($p<0.1$) advantage for the 2D U-Net of 37\% to 75\%.

\begin{table*}[!hbt]
\centering
\begin{tabular}{l||l|l||l|l}
\multicolumn{1}{l||}{\bfseries DSCs:} & \multicolumn{2}{|l||}{\bfseries Mean\(\pm\)Std.} & \multicolumn{2}{|l}{\bfseries Median\(\sim\)IQR}\\
Organ & 2D U-Net & 3D U-Net & 2D U-Net & 3D U-Net\\
\hline
\hline Liver 
& \textbf{0.94\(\pm\)0.03\textsuperscript{\textbf{\footnotesize*}}}
& 0.93\(\pm\)0.04
& \textbf{0.95\(\sim\)0.02\textsuperscript{\textbf{\footnotesize**}}}
& 0.94\(\sim\)0.03
\\
\hline R. kidney 
& \textbf{0.91\(\pm\)0.05\textsuperscript{\textbf{\footnotesize***}}}
& 0.89\(\pm\)0.05
& \textbf{0.92\(\sim\)0.03\textsuperscript{\textbf{\footnotesize***}}}
& 0.90\(\sim\)0.05
\\
\hline L. kidney 
& \textbf{0.92\(\pm\)0.05\textsuperscript{\textbf{\footnotesize***}}}
& 0.86\(\pm\)0.14
& \textbf{0.93\(\sim\)0.03\textsuperscript{\textbf{\footnotesize***}}}
& 0.89\(\sim\)0.08
\\
\hline Spleen
& \textbf{0.93\(\pm\)0.04}
& 0.92\(\pm\)0.04
& \textbf{0.94\(\sim\)0.03}
& 0.93\(\sim\)0.03
\\
\hline Pancreas 
& 0.57\(\pm\)0.19
& \textbf{0.59\(\pm\)0.15}
& 0.60\(\sim\)0.21
& \textbf{0.61\(\sim\)0.21}
\end{tabular}
  \caption{Mean DSCs with standard deviations (Mean$\pm$Std.) and Median DSCs with Inter-Quartile-Range (IQR) (Median$\sim$IQR) of 2D and 3D U-Nets from 5-fold randomized cross-validation experiments using 4:1 splits of the 80 images into training and test data.\\\textbf{Legend}: We use statistical standard notation for found significances: \textbf{*}; \textbf{**}; \textbf{***}: $p<0.05$; $p<0.01$; $p<0.001$ from T-tests (paired) and Wilcoxon-Signed-Rank-Tests on the right of the \textbf{favorable U-Net result}.}%
  \label{tab:dice}
\end{table*}

\begin{table*}[!hbt]
\centering
\resizebox{\textwidth}{!}{%
\begin{tabular}{l||l|l|l|l||l|l|l|l}
\multicolumn{1}{l||}{\bfseries GPU-Performance:} & \multicolumn{4}{|l||}{\bfseries Memory} & \multicolumn{4}{|l}{\bfseries Time} \\
\multicolumn{1}{l||}{} & \multicolumn{2}{|l|}{\bfseries Training [MiB]} & \multicolumn{2}{|l||}{\bfseries Application [MiB]} & \multicolumn{2}{|l|}{\bfseries Training [min:sec]} & \multicolumn{2}{|l}{\bfseries Application [sec]}\\
Organ & 2D U-Net & 3D U-Net & 2D U-Net & 3D U-Net & 2D U-Net & 3D U-Net & 2D U-Net & 3D U-Net\\
\hline
\hline Liver 
& 1693
& 10957
& 1693
& 9117
& 9:26
& 10:07
& 1.47
& 3.18
\\
\hline R. kidney 
& 1693
& 10957
& 1693
& 9117
& 9:28
& 10:07
& 0.40
& 0.55
\\
\hline L. kidney 
& 1693
& 10957
& 1693
& 9117
& 9:26
& 10:07
& 0.40
& 0.55
\\
\hline Spleen
& 1693
& 10957
& 1693
& 9117
& 9:27
& 10:08
& 0.40
& 0.55
\\
\hline Pancreas 
& 1693
& 10957
& 1693
& 9117
& 9:28
& 10:08
& 0.40
& 0.55
\\
\hline
\hline Mean\(\pm\)Std. 
& \textbf{1693\(\pm\)0\textsuperscript{\textbf{\footnotesize***}}}
& 10957\(\pm\)0
& \textbf{1693\(\pm\)0\textsuperscript{\textbf{\footnotesize***}}}
& 9117\(\pm\)0
& \textbf{9:27\(\pm\)0:01\textsuperscript{\textbf{\footnotesize***}}}
& 10:07\(\pm\)0:01
& \textbf{0.61\(\pm\)0.43}
& 1.07\(\pm\)1.05
\\
\hline Median\(\sim\)IQR 
& \textbf{1693\(\sim\)0\textsuperscript{\textbf{\footnotesize***}}}
& 10957\(\sim\)0
& \textbf{1693\(\sim\)0\textsuperscript{\textbf{\footnotesize***}}}
& 9117\(\sim\)0
& \textbf{9:27\(\sim\)0:01\textsuperscript{\textbf{\footnotesize+}}}
& 10:07\(\sim\)0:01
& \textbf{0.40\(\sim\)0.003\textsuperscript{\textbf{\footnotesize+}}}
& 0.55\(\sim\)0.001
\\
\hline Mean Improvement 
& \textbf{647.19\%}
& N/A
& \textbf{538.51\%}
& N/A
& \textbf{107.13\%}
& N/A
& \textbf{175.22\%}
& N/A
\\
\hline Median Improvement 
& \textbf{647.19\%}
& N/A
& \textbf{538.51\%}
& N/A
& \textbf{107.07\%}
& N/A
& \textbf{137.10\%}
& N/A
\end{tabular}
}
  \caption{GPU-Performance table with memory consumption for training and application to the left and training and application times to the right.\\\textbf{Legend}: 1 MiB=1.048581024 MB=1024\textsuperscript{2} bytes. We use statistical standard notation for found significances: \textbf{$+$}; \textbf{*}; \textbf{**}; \textbf{***}: $p<0.10$; $p<0.05$; $p<0.01$; $p<0.001$ from T-tests (paired) and Wilcoxon-Signed-Rank-Tests on the right of the \textbf{favorable U-Net result}.}
  \label{tab:GPU-performance}
\end{table*}

\section{Conclusion}

Surprisingly in this study, 2D U-Nets are favorable regarding the ratio of quality vs. computing costs.

The liver and spleen hold the greatest volume in our abdominal organ group. The liver and especially kidney competition is significantly won by the 2D U-Net ($p<0.05$). The liver is a difficult organ often with a variably filled stomach as a neighbor. 

Kidneys can be regarded as easy organs lighted by contrast agent and inside fatty tissue with low CT intensity.

A better posed training for the 2D U-Net could be the reason for the 2D U-Nets' better results for liver and kidney tasks. A higher relative number training elements is used, i.e. axial slice pixels, vs. the number of net weights. 

The spleen results are in favor of the 2D U-Net regarding the mean and medians by trend. The 2D U-Net is also favorable for a less number of outliers (Fig. \ref{plot:boxplotOverviewSpleen}). 

The difficult pancreas does not provide many axial training slices useful for the 2D U-Net, as its elongation is not prominent on the z-axis. We suppose this is the reason, why the 3D U-Net wins here by mean trend in terms of accuracy and precision, i.e. higher mean and lower standard deviation and lower number of outliers (Fig. \ref{plot:boxplotOverviewPancreas}). This win is supported by higher accuracy for the 3D U-Net in terms of medians. However, the race is not decided by significant differences making the 2D U-Net still very attractive for some users with GPU performance and memory concerns.

Abdominal volumetric CT images and key organ segmentation were analyzed.
This new study shows interesting results from competing U-Net architectures, especially focusing different dimensionalities of net filter bank kernels and quality vs. GPU performance. The interested reader can now select a particular U-Net architecture, primarily whether to use a computational inexpensive design. Finally in this study's scope, a humble recommendation for the 3D U-Net could be given for the pancreatic organ in terms of better accuracy by trend and smaller standard deviation only. The IQRs as an alternative measure of precision are on par, and regarding the median the 3D U-Net wins just by a small trend. We co-conclude, because of the deeper layering structure and thus more trainable weights, a 3D U-Net needs significantly more training data vs. possible overfitting to outpace 2D U-Nets. 

As training volumes are normalized to a square or cube of 96 voxels, the GPU memory consumption in Tab. \ref{tab:GPU-performance} is always constant. Therefore differences are trivially highly significant, as no varying results occur. We observe consistently lower memory consumption for 2D U-Nets. The memory effort in training is higher for the 3D U-Net including more space for administrative overhead data. Thus, 2D U-Nets can run on affordable 2-4 GB GPUs for 3D CT volume segmentation. 3D U-Nets definitely need currently totally overpriced 12 GB GPUs.

The timing measures in Tab. \ref{tab:GPU-performance} clearly speak out for the 2D U-Net. Training is highly significantly 40 seconds or 7\% faster on average using the 2D U-Net. However regarding application, in the trained model prediction, the differences are not so striking with a weak significance by median. 37\% to 75\% improvement can be achieved, however in the range of one second, which is practically unimportant. 
As final and bold conclusion regarding our study design and results, we can recommend using the 3D U-Net subject to the amount of data we used here - for pancreas only. The conclusions are justified by statistically significant or by trend quality and GPU-computation performance results for all organs under study. We suppose, 3D U-Nets may overcome in quality when using several hundreds of training images. However, this comes approximately with an order of magnitude higher additional computational burden.

For the first time,
an original and significant comparison of U-Net architecture dimension is provided to the reader focusing the key abdominal organs of liver, spleen, kidneys and pancreas. The reader can decide, which approach is appropriate for his concrete target organ, amount of training data and used GPU or cluster nodes, parallel process design, e.g. for atlas-based usage of U-Nets as encountered in multi-classifier fusion\cite{meine2018comparison}.

Regarding the difficult pancreas with mixed DSC results in this study, we plan to train 2D U-Nets using a elongation optimized algorithm\cite{reinbacher2010variational,kallergi2005automatic,mastmeyer2015model} to provide slices oriented axially along its main central curve, to better reflect its orientation to generate more training slices for 2D U-Nets. On the other hand, more training images shall be used to explore the 3D U-Nets' theoretical advantage under better conditions, as we have discussed, for its training and application.

The scientific explanation of the U-Net methods is given detailed in the original works of Ronneberger and Cicek et al.\cite{Cicek16,Ronneberger15}, and we lift these methods here to compare them. 
The current limitations of this study will manifest as improvements in the future. Development will cover improved bounding box detection. At this state of our research, our current RRF bounding box detection would confound the core message of this paper, the aim of which was to focus purely on U-Net performances.

\section{Acknowledgments}
German Research Foundation: DFG MA 6791/1-1;\\
Nvidia GPU grant;\\
Foundation Kessler+Co. for Education and Research: EXPLOR19-AM.
\printbibliography
\end{document}